\documentstyle[epsfig,12pt]{article}
\begin{document}

\vspace{2cm}
\begin{center}
{\Large {\bf The Physics of ALICE HLT Trigger Modes}}

\vspace{0.5cm}
 R.~Bramm, T.~Kollegger, C.~Loizides and R.~Stock,

\{bramm,kollegge,loizides,stock\}@ikf.uni-frankfurt.de,

Physics Department, University of Frankfurt/Main
\end{center}

\abstract{We discuss different physics cases, mainly of the ALICE TPC, such as pile-up, jets in pp and PbPb, Bottonium and Charmonium spectroscopy, and there corresponding demands on the ALICE High Level Trigger (HLT) System. We show that compression and filter strategies can reduce the data volume by factors of 5 to 10. By reconstructing (sub)events with the HLT, background events can be rejected with a factor of up to 100 while keeping the signal (low cross-section probes). Altogether the HLT improves the discussed physics capabilities of ALICE by a factor of 5-100 in terms of statistics.}

\section{Physics Motivation}

The Physics of ALICE can  be broadly grouped into two types of
physics observables, which connect to the soft and hard sectors of
QCD, respectively. Among the former group one finds the physics of
bulk hadron production, multihadron correlations and
interferometry, fluctuations of short and long ranges, resonance
and string decay signals, as well as multistrange hyperons. Mostly these
observables refer to aspects of nonperturbative QCD even if their
initialization makes contact to the primordial partonic
"transport" dynamics in which (at $\sqrt{s}$ = 5.5 TeV per
participant nucleon pair in PbPb collisions) perturbative QCD
obvious plays a role, too. Pertubative QCD (pQCD) aspects dominate, however, in the
second group of observables consisting of various aspects of jet
and heavy flavour production. It is the main point of such studies
to connect the initial dynamics during interpenetration of the
interacting primordial hadronic ground state matter (supposed to
be well understood from modern pQCD application to elementary
$p\overline{p}$ collisions) to the ensuing parton cascade era. In
that era the initial pQCD "seedlings" such as $b\overline{b}$
pairs or very energetic quarks and gluons may act as "tracers" of
a dynamical phase that may approach the deconfined quark-gluon
state envisaged in non perturbative Lattice QCD theory, before it
hadronizes. The resulting final signals, e.g. jets, open charm and
quarkonia should thus exhibit certain characteristic attenuation
properties (suppressions, enhancements, quenching etc.) that might
serve as diagnostic tools for the intermediate, non perturbative
QCD era of near equilibrium partonic matter.
It is clear from the above that all such observables also require
systematic study of the more elementary $pp$ collisions at LHC
energy, in order to establish the base line physics - a further
research focus of the ALICE experiment.

Clearly there is a coincidence of the "mostly soft" and
"predominantly hard" QCD physics sectors of ALICE with relatively
large, and small cross sections, respectively. It turns out that
analysis with the former type of physics should require relatively
modest event statistics, of a few $10^6$ PbPb and about $10^8$
pp collisions, whereas the systematic analysis of hard signals
calls for an additional one or two orders of magnitude both in
PbPb and pp. For example, inclusive production of jets with
total transversal energy of up to 200 GeV, or of the weaker states
in the bottonium family, is expected to occur (within the ALICE
tracking acceptance) about once every $10^4$ to 10$^5$ central
PbPb collisions. It is the latter sector of ALICE physics which
will be addressed below. Obviously some kind of on-line higher
level trigger selectivity is being called for, in order not to be
drowned in excessive demands placed both on DAQ bandwidth and
off-line data handling format.

More specifically, ALICE differs from the other LHC experiments by
its insistence on microscopic track by track analysis of the
events, both in the central barrel (tracking from the ITS through
the TPC and the TRD into the outer TOF layer and the photon array)
and in the dimuon spectrometer. Calorimeters with a more summative
output are absent in this basic design although an augmentation,
by an electromagnetic calorimeter, remains as an upgrade option of
special interest for jet spectroscopy. The consequence of the
insistence on single track resolution leads to a raw data flux
(after on-line zero suppression and Huffman compression in
the digital front end electronics) amounting to about 40-80
Megabytes in a central PbPb collision, which presents us with up
to about 20.000 charged tracks per event. Clearly, high statistics
demands as associated with the above small cross section signals
meet with excessive raw data volume calling for higher on-line
selectivity based on track pattern recognition.

The required on-line selectivity can be accomplished with relatively 
modest computing and connectivity power, using an High Level Trigger 
(HLT) complex of about 1000 CPUs equipped with fast interconnectivity. 
A major part of the required processor power 
can be implemented already in the
local data concentrator circuits (LDC) of the DAQ front end
instrumentation that is devoted to the raw data flux from the
ALICE barrel TPC detector which produces, by almost an order of
magnitude, the predominating data input on the ALICE DAQ system.
We shall illustrate below how the rarest physics signals
(cross-section wise) provide for the most topologically distinct
tracking signatures in the TPC subdetector - recognizable with
relatively modest CPU-networking effort. We thus describe the
physics cases motivating on-line HLT trigger schemes based on
on-line tracking of the ALICE TPC output. Further refinement may
result from calling in early time information of the ITS and TRD
systems. The dimuon-forward spectrometer works outside the
acceptance of the barrel detectors. Its HLT requires separate
consideration but is of minor format as compared to the TPC HLT.
The following sections describe the raw outline of an HLT system
that will cope with an on-line degradation of a TPC event rate of
200 Hz for central PbPb collisions. There are three basic modes
to such an HLT functionality. It may either generate selected
candidate events according to the detected track topologies
corresponding to certain specific trigger tasks. We will focus on
jet spectroscopy at first, with a background suppression
efficiency of between 50 and 100: thus a 200 Hz TPC event rate
will generate 2 to 4 candidate events per second, which are
written onto tape in full raw data format. Alternatively, the HLT
system might reduce the full TPC sector readout to
region-of-interest readout. E.g. reading out  2 sectors only (out
of 36) in conjunction with a TRD-generated pre-candidate for a
near back-to-back $e^+e^-$ track pair resulting from bottonium
decay will lead to an 18-fold compression of data flux. Finally,
in later years of ALICE operation one may consider the option of
employing the TPC HLT circuitry only insofar as on-line cluster
identification is concerned, which then would be recorded by DAQ
instead of the full raw data arrays leading to an overall TPC data
compactification by about 10 to 20. The latter mode will be
employed in the next years by the RHIC experiment STAR, which
resembles ALICE in layout. In summary there are three HLT modes:
selective trigger, region of interest readout (filter) and global
TPC raw data compression.

We have constructed and tested an HLT on-line tracking prototype
at the RHIC STAR experiment (called L3 there) with 54 processors
confronting the STAR TPC data format, which is about 10\% of the
expected event size for ALICE. Both the selective trigger and
overall compactification (by a factor of 6 in this prototype case)
modes work well. In particular, the on-line momentum and specific
ionization resolutions of that device operating at a maximum rate
of 50 Hz for central collisions of Au+Au at RHIC was shown to
approach the off-line resolutions within about 20\%. In later
sections we shall extrapolate from this L3 to the ALICE HLT to
obtain a guide line concerning ALICE HLT processing power demands.

In the following sections we shall at first specify the ALICE
running conditions in terms of LHC luminosity with special
emphasis on the TPC event rate, which is the slowest ALICE
subdetector but creates, by far, the highest raw data flux.
However, it also creates conditions of several superimposed events
captured in one TPC output frame. That condition can be dealt with
by selection of "clean events" in PbPb collisions (by lower level
triggers) but only by an HLT on-line filter procedure in pp
collisions at higher luminosity. This will lead to the proposal
that data taking with inclusion of the TPC (there are other
options in ALICE) can proceed at 200 Hz in central PbPb, and at 1
KHz in pp. With these options we arrive at an optimum TPC
operation vis a vis the LHC luminosities.

Both options would create an event rate that is necessary for jet
and bottonium spectroscopy, but can not be written directly to
tape, because of the resulting raw data flux of about 10 to 15
Gbyte per second from the TPC alone. Further sections thus deal
with the required expected HLT speed and selectivity in various
modes. For illustration we shall mostly focus on high transverse
energy jet spectroscopy, both in PbPb and in pp collisions. A
consideration of other HLT trigger modes ends these sections of
the document.

\subsection{Running Conditions of ALICE}

At the design luminosities concerning the ALICE experiment we
encounter high interaction rates which are fully satisfactory for
all observables of concern. However 8 KHz minimum bias PbPb rate
as well as pp at 140 KHz would both create roughly 0.5 Terrabytes
of potential raw data per second which we can not write to mass
storage. This estimate already takes into account on-line zero
suppression in all its high granularity subdetectors. Any kind of
on-line intelligence is thus welcome to

\begin{enumerate}
\item maintain event rates appropriate for low cross section
physics signals (charm and bottom spectroscopy, high $E_T$ jets),
which range down to one in $10^7$ pp or to one in $10^4$ PbPb events

and to
\item reduce the bandwidths of DAQ processing and writing
to tape, and the magnitude of data deposited and handled in mass
storage. This is important as neither the taping speed has
increased significantly (in a fashion resembling other component's
"Moore's law" pattern), nor has the tape and robot media cost
shown a marked decrease. One is eager, therefore, to keep these
cost factors within reasonable bounds, while increasing the
overall physics content of the raw data flow.
\end{enumerate}

In view of point 2 it would be desirable to cut down to say 1.0
GB/s taping rate, i.e.~to about 0.5\% of the raw data rate
corresponding to the luminosity limits. In order to achieve  this
we must employ any suitable method of front-end/DAQ intelligence,
in particular with the data-intensive subdetectors TPC, TRD and
Dimuon spectrometer. Such methods start with zero suppression and
automatic, loss free data compactification (Huffman etc.), the
latter  expected to reach a compression by about two. These
methods will not concern us here. Instead we 
concentrate on high level trigger- and filter-procedures (HLT)
which are based on on-line tracking in large
processor/switch/storage arrays placed at the front-end of the
DAQ. We will comment on the physics driving such HLT-efforts,
primarily, because we need first to assess the potential physics
benefit, in order to conclude about the appropriate efforts in HLT
and DAQ. The first topics that we choose to illustrate the
expected HLT will be
\begin{enumerate}
\item Jet physics in pp ("first year" and beyond)
\item Jet physics in PbPb
\item Y spectroscopy in TRD and TPC, and
\item Event filter procedures for TPC/ITS open charm analysis
\end{enumerate}
Some of these topics concern genuine triggers enabling us to scan upward of 200 Hz of
PbPb and up to one KHz of pp, without exceeding a reasonable DAQ
writing budget. In other cases the filter aspect dominates:
recording the accessible physics with lower data volumes.

\subsubsection{Technical Assumptions}

At average luminosity of $10^{27}$ the minimum bias rate (assuming
8 barn total cross section) for PbPb is 8 KHz, of which we may
label 1 KHz as "central". The bunch spacing is 100 ns, thus we get
$1.2  \cdot 10^{-3}$ events per bunch crossing, i.e.~very rarely
we indeed get two events within the crossing time of about 100 ps.
For KrKr at $5 \cdot 10^{28}$ luminosity and 4 barn the min.~bias
rate is 240 KHz, and the fraction of events unresolvable to our
electronics  rises to the 1 \% level.

In pp we will have full LHC energy at first of $\sqrt{s}$=14 TeV
(unlike in PbPb). We assume a luminosity of $2 \cdot 10^{30}$
here, which would lead to 140 KHz minimum bias rate for a 70 mbarn
non-diffractive cross section. The above luminosity is already
non-trivially low for the LHC overall running conditions (in
combination with the maximal luminosities employed in the other
experiments it would be simpler to run at far higher L). A lower
luminosity for ALICE will be accomplished either by a different
$\beta^*$ or by de-tuning (or both). One concludes that we can not
safely assume that the 2~$\sigma$ size of the interaction domain
(the "diamond") will initially be as low as 10~cm. Furthermore we
will argue below that a further reduction of pp luminosity to the
$10^{29}$ level is highly desirable: a challenge to LHC technique.

\subsubsection{TPC Event Pileup and Rate}

At the above event rates the TPC (90 $\mu$s drift time) has a
significant double event fraction within the drift time already
for PbPb which gets forbiddingly high for KrKr and CaCa at higher
potential luminosities (no TPC physics possible). In pp at L=$2
\cdot 10^{30}$ we will have many (but smaller) events in each
frame, displaced by the drift velocity of 2.8 cm/$\mu$s. More
precisely the average fraction of PbPb double events in the TPC at
8 KHz minimum bias rate is [1-exp(-2$\tau_{drift}
 \cdot f)] =0.76$ where we have taken 2$\tau$ because
 each TPC frame contains displaced tracks/events occurring both
 90~$\mu$s before and after the trigger. This effect is specific to
 the TPC as all other subdetectors have drift - or integration
 times of up to 5 $\mu$s only. The "clean" minimum bias PbPb
 event rate thus shrinks to 1.900 Hz, the central rate to  240 Hz
 as far as the TPC is concerned; both at L (average)=10$^{27}$.
  This limitation of useful event rate roughly coincides with two other
 TPC limitations, the maximum possible TPC gating frequency (now
 estimated to be about 1 KHz), and the maximum data
 transfer rate from the TPC front-end electronics to the HLT/DAQ
 complex as implied by the 216 DDL links, whose individual bandwidth is  assumed here
 to amount to 150 MB/s. Estimating a "clean" central PbPb event to
 contain 66 MB after zero suppression and a flat
 distribution of data traffic over all the TPC-DDLs we would get a maximum
 event transmission rate of 400/s. However, we already get into double events with
 about
 100 MB event size this way, and, furthermore, the traffic
 distribution may not be flat over all DDLs. Thus we take
 about 200~Hz as a, perhaps preliminary, technical limit for "clean"
 central PbPb collisions and about 400~Hz for min.~bias for any
 ALICE data taking mode involving the TPC. The HLT system will
 thus be presented in a version that copes with the corresponding
 raw data input of up to about 15-20 GB per second in PbPb collisions.

\subsubsection{TPC Pileup in PP}

In pp running at L=$2 \cdot 10^{30}$ and $\sqrt{s}=14$ TeV the TPC
physics gets quite challenging \cite{Giubellino}. At 70 mb we get 140 KHz
min.~bias event rate. From prior and post 90 $\mu$s to the trigger
25 events fall into one TPC frame which are on average
half-complete. Average spacing in time is about 7 $\mu$s but note
that the events are not therefore ordered in drift distance
 because of the variation of the primary
vertex position which will randomize distances, perhaps even
inverting the ordering. This influences the capability to identify
the sub-event that belongs to the trigger, by means of on-line
tracking in the HLT leading to an output to DAQ that contains the
relevant sub-event only (HLT filter mode).

The  TPC data flux per event arising
 in a situation with 25 half-complete min.~bias pp
collisions at $\sqrt{s}=14$ TeV is estimated to be about 4.5~MB;
making allowance for a certain non-track-density-related noise
(e.g. electronics) we conservatively estimate 5-6 MB, i.e.~roughly
8\% of a central PbPb event.

The maximum estimated TPC gating frequency of 1 KHz would thus
result in a front-end data output rate of about 5.5 GB/s,
equivalent to the output of 85 Hz worth of central PbPb. From the
above considerations this is well compatible with the DDL and HLT
bandwidths but can not be written to tape. An HLT facility is thus
required to filter out the relevant information contained in each
piled-up TPC frame.

\subsubsection{Summary of ALICE Running Conditions}

Consideration of luminosity, event pileup conditions, maximum TPC
gating rate, as well as data rate capability of the TPC front-end
electronics including the DDL bandwidth leads to an estimate of
maximum event frequencies as far as the TPC is concerned:

\vspace{0.6cm}
\begin{center}
\fbox{\parbox{9cm}{ 1 KHz for minimum bias pp collisions

400 Hz for pileup-free PbPb min.~bias collisions

200 Hz for pileup-free PbPb central collisions}}
\end{center}
\vspace{0.3cm}

\noindent The other ALICE subdetectors can run along with these
event frequencies, but can not accept significantly higher event
rates anyhow - perhaps going up to 1.5 KHz for min.~bias and
central PbPb collisions in the ITS, TRD and Dimuon detectors, thus
capable of defining detector specific level 3 pre-triggers for a
pre-scaling that involves the TPC at its lower appropriate rates.

\vspace{0.2cm} \noindent However,

\vspace{0.3cm}
\begin{center}
\fbox{\parbox{11cm}{ 1 KHz pp collisions yield about 5.5 GB/s
front-end flow

400 Hz min.~bias yield about 15 GB/s front-end flow

200 Hz central PbPb yield about15-20 GB/s front-end flow}}
\end{center}
\vspace{0.3cm}

\noindent from the TPC alone. We shall show below that the
intended ALICE physics requires such rates, which we can not write
 to tape. This consideration makes the case for an intelligent
processor front-end system attached to the DAQ, deriving specific
triggers and/or reducing the raw event sizes by appropriate
filtering procedures: the High Level Trigger (HLT) system. In the
following we will first of all consider the task of jet
spectroscopy in pp and PbPb. These physics observables will serve
to illustrate, both the minimal (pp) and maximal (PbPb)
requirements, placed on the HLT functions. A short sketch of other
physics observables concludes these considerations.

\subsection{Jet Physics in pp and PbPb Collisions}

High transverse energy jet production at LHC energy is of interest
both from the point of view of higher order QCD (e.g. twist and
gluon saturation) \cite{Mangano}  and from the program of
quantifying jet attenuation in extended partonic matter, i.e.~in
nuclear collisions at ALICE. The overall goal of the latter idea
is to determine the QCD "stopping power" acting on a colour charge
traversing a medium of colour charges, in analogy to the
Bethe-Bloch physics of QED. This effect will attenuate the energy
of the observed jets. Predictions of perturbative QCD as to the
mechanism of energetic parton propagation in hot and cold QCD
matter address the borderline of present state of the art theory,
thus recei\-ving increasing attention. For a length L traversed in
QCD matter the induced radiative energy loss is proportional to
L$^2$ and expected to be much higher in a parton plasma than in
colder hadronic matter even at moderate plasma temperatures of
T$\approx$200 MeV. The effect of "jet quenching" can be estimated
\cite{Baier}  to amount to

\vspace{0.2cm}
\begin{math}
-\Delta E \approx 60\: GeV (\frac{L}{10 fm})^2 \approx 30 \:GeV
for \:L=7.5 fm.
\end{math}

\vspace{0.2cm}

\noindent Here, L is the transverse radius of the PbPb
"fire-sausage", which is the relevant geome\-tri\-cal length scale
as we are observing jets with average center of mass angle of 90$^0$ with
the ALICE TPC positioned at $\mid \eta_{CM} \mid < \:1$. Thus the
main task is to measure the inclusive jet production yield at
mid-rapidity as a function of E$_T$ both in pp and central PbPb.
Of course one is, more specifically, interested in the details of
the jet fragmentation function \cite{Affolder}. The capability of
ALICE to reconstruct in detail the jet fragmentation function in
the charged track sector is crucially important in view of ongoing
theoretical study concerning the detailed consequences of
primordial partonic mechanisms of energy loss. It turns out that
the hadronization outcome, from the energy loss components of the
leading jet parton, may also be partially contained within the
typical jet emission cone, resulting in an increase of local E$_T$
emission but expressed by relatively softer hadrons.

The picture of quantifying energy loss of the primordial jet may
thus be an over-simplification. One may have to investigate,
instead, the microscopic changes, at track-by-track level,
occuring within the fragmentation cone of a high E$_T$ jet created
in PbPb collisions, in order to appropriately capture the overall
attenuation effect acting on a leading parton traversing high
energy density partonic matter sections of the interaction
"fireball". In this view, the jet attenuation phenomena should
reside in a characteristic softening of the jet cone fragmentation
function, rather than in a simple jet energy loss scenario, as
considered above. The microscopic tracking approach of ALICE may
thus prove advantageous, in comparison to the summative E$_T$
inspection by calorimeters.

\subsubsection {HLT Function in PP for Low Cross Section Signals}

\noindent As stated above, we probably will get only $\sqrt{s}=14$ TeV initially
for the pp beam. For low cross sections signals such as jets, $\Omega$
and Y we need about 10$^{10}$ events, in order to record

\hspace{4,5 cm}  5 $\cdot$ 10$^4$ jets with E$_T \ge$ 100 GeV

\hspace{4,5 cm} 10$^5$ $\Omega$ in acceptance

\hspace{4,5 cm} 10$^3$ Y $\rightarrow e^+e^-$ in acceptance.

The ideal solution would be to reduce the luminosity further
(below 2 $\cdot10^{30}$), which technically is very difficult.
Assuming a luminosity of 10$^{29}$ would result in an average of
1.7 min.~bias events per TPC frame at 7 kHz event rate. For that scenario
an average TPC event content of $\approx$ 0.5 MB at a maximum trigger rate
of 1000 Hz will lead to 0.5 GB/s to DAQ. This is a writing speed compatible
with ''first year'' DAQ without employing an HLT.

\noindent If LHC can not go lower than a luminosity of 2 $\cdot
10^{30}$ at ALICE for technical reasons, we need to employ HLT
functionality. Now the average number of events per TPC frame is
about 25 partially half events pile-up and thus the average
TPC event content is roughly 5 MB. Therefore at 100 Hz DAQ
saturation occurs assuming a first year 0.5 GB/s bandwidth and we
would get only 10$^9$ events per year (i.e.~5 months), which is
unsatisfactory.

Employing HLT for event filtering at full TPC frequency of 1000 Hz
to filter out the trigger event only, we expect at least a 10:1
reduction of raw data flux. Therefore at the same DAQ bandwidth of
0.5 GB/s to tape we could store 10 times the amount of events thus
exhausting the maximal TPC gating frequency. In that scenario
HLT-processors for about 250.000 tracks/s are required. Derived
DAQ bandwidth and data storage needs are

\vspace{0.3cm}

\begin{center}
\fbox{\parbox{10cm}{
\hspace{0.35cm}0.5 GB/s to tape from TPC alone for the first year.

\hspace{0.20cm} 0.7 GB/s to tape from TPC, TRD and ITS together.
}}
\end{center}

\noindent That leeds to a total storage of 5 Petabyte per year pp, which
is very high, but not done every year.

\vspace{0.3cm} \noindent In the following section we shall
demonstrate that the typical one-year data output of ALICE for
PbPb running 2007 and beyond will fall in the vicinity of 1.8
Petabyte "only". However, due to the relatively short expected Pb
beam periods of LHC, the DAQ bandwidth has to increase from about
0.5 GB/s in first year to about 1-1.2 GB/s after 2007, in spite of
increasing HLT effort.

\subsubsection{Jet Physics in PbPb Collisions with HLT Trigger}

The physics questions to be persued first are: How does the
inclusive jet cross section at or near 90$^0$ vary with the total
transverse energy E$_T$? And what are the properties of the
fragmentation function? Both to be compared to elementary data
from pp.

Recall M.~Mangano's estimate \cite{Mangano} of one jet with E$_T
\ge$ 100 GeV in $2 \cdot 10^{5}$ events for the reduced TPC
acceptance of $\mid \eta \mid < 0.5$ at $\sqrt{s}$=14 TeV
min.~bias pp collisions. How to scale to central PbPb collisions
at $\sqrt{s}$=5.5 TeV? Employ QCD scaling to extrapolate the jet
cross section from 630 and 1800 GeV $\sqrt{s}$ as reported by D0
and CDF \cite{Affolder}. As to the dependence of jet multiplicity
on target/projectile participant nucleon pair number in AA the QCD
expectation is A$^{4/3}$ where A$\approx$185 in a "central" PbPb
collision. Note the extra factor of 1/3 in the exponent over
"wounded nucleon" models that scale with A. Thus the jet
multiplicity for standard physics would increase by A$^{4/3}$ but
decrease by about 10 due to energy scaling down from $\sqrt{s}$=
14 TeV (Mangano estimate) to $\sqrt{s}$=5.5 TeV. This leaves a
gain factor of about 100 from pp at $\sqrt{s}$=14 TeV to AA at
$\sqrt{s}$=5.5 TeV. The interesting first order effect of "jet
attenuation" degrades the jet energy by the radiative loss of
$-\Delta$E$_T$. Baier et al. \cite{Baier} estimate this to be -30
GeV at $\Theta \approx$90$^0$ in PbPb (see the previous section).
This shift, if independent of E$_T$, would reduce the jet
multiplicity by a factor of about four, for the case of 130~GeV
jets shifted down to an observed 100 GeV in central PbPb. This
reduces the expected overall scaling factor from pp to PbPb to a
final value of 25, on which we will base the considerations below:

\vspace{0.6cm}
\begin{center}
\fbox{\parbox{11cm}{ The jet multiplicity per central PbPb at
$\sqrt{s}$ =5.5 TeV equals that of 25 min.~bias pp collisions at
$\sqrt{s}$=14 TeV.

\noindent This results in one E$_T\ge$ 100 GeV jet every about
9000 central PbPb, in the effective TPC acceptance (for full jet
containment) of $\mid \eta \mid < 1/2$.}}
\end{center}

\vspace{0.6cm} \noindent Note that we are focusing here on jets at
and above 100 GeV in order to stay compatible with the assumption
 that jets can be identified by relatively simple on-line
tracking algorithms. At E$_T\ge 100$ GeV jets have on average a
unique charged track topology and a sufficient charged track
multiplicity of $<n_{ch}> \approx 10$, to stand out over the
fluctuating mini-jet background even in central PbPb collisions.

\noindent {\bf Jets at 100 $\le$ E$_T \le$ 200 GeV in central
PbPb, $\sqrt{s}=$ 5.5 TeV}

\noindent We assume again E$_T$ attenuation by perhaps up to 30
GeV. There are around 10 charged particles in the jet. The
"leading particles" of a 100 GeV jet are in the 20$<$E$_T<$30 ~GeV
domain, which is compatible with TPC E$_T$ tracking resolution.

\vspace{0.5 cm} \noindent In such a scenario we need more than
about 10$^8$ inspected events in the TPC: this will give $\ge
10^4$ events with E$_T \ge 100$ GeV even at 30 GeV energy
attenuation.

\noindent One PbPb year corresponds to 10$^6$ s or 2 $\cdot 10^6$
s run time with an inspection rate of 200 Hz in TPC, which we can
not write to tape for obvious reason. But employing an HLT based
on on-line jet-finder tracking algorithms would result in 200 Hz
central PbPb TPC inspection by the HLT, which yields $\ge $ 50
jets per hour with attenuation. The resulting maximum HLT duty is
to reconstruct up to 2 $\cdot 10^6$ tracks/s.

We will sketch below an HLT TPC-tracking-based jet finder
algorithm that could serve as a jet-trigger for E$_T \ge $100 GeV
jets, capable of a 200 Hz central PbPb collision rate. Before
dealing in more detail with HLT jet triggering we sketch other
ALICE HLT tasks. However, let us conclude here that both the
minimum demands (pile-up in pp) and the maximum demands (jets in
PbPb) on the HLT functionality have been involved above for jet
spectroscopy. All further observables will stay well within this
HLT budget.

\subsection {Bottonium Spectroscopy}


We consider bottonium spectroscopy and chose as a key example the
decay of $\Upsilon$ (9460) $\rightarrow e^+e^-$, which creates two electron
tracks of about 5 GeV. The TRD is constructed to increase electron
selectivity over pions of equal momentum. The HLT-trigger is thus
built on TRD precognition of an $e^+e^-$ pair (with roughly
correct invariant mass). This pre-trigger hands down two "regions
of interest" (ROI) to the TPC, thus involving inspection of two
TPC-sectors for more accurate verification and momentum plus
specific ionization measurement of the two tracks, leading to fast
rejection or invariant mass verification on-line. The latter
activity carried out by the TPC HLT processors leads to an $\Upsilon$
candidate trigger. The $\Upsilon$ cross section is as low as the jet
physics cross section considered for ALICE thus far. We estimate
one $\Upsilon\rightarrow e^+e^-$ decay in the TPC/TRD acceptance in about
$10^4$ central PbPb collisions (including electron pair efficiency
in the TRD and dead areas). This estimate is based on a full
coverage of the TPC by the TRD. Full TRD coverage represents a
future ALICE upgrade potential. With about 60\% coverage in
initial ALICE running, it thus follows that about $10^8$ PbPb
collisions are required to accomplish a satisfactory $\Upsilon$
spectroscopy. Without HLT functions the maximum yearly ALICE event
statistics would amount to about $10^7$ events only.

In order to discuss ways of proceeding to higher event frequencies
than 10-20 Hz it is important first to note the technical
constraints: the TRD, by itself, could perhaps trigger at 500 Hz,
but if we insist that any $\Upsilon$ candidate event preselected by the TRD
meets with a concurrent pile-up free TPC event, we have to cut
down to 200 Hz for central PbPb and to 400 Hz for min.~bias PbPb
due to the TPC limitations. Both running modes will yield about
the same number of $\Upsilon$ decays per unit time. Let us, therefore,
concentrate on central collisions. At a common HLT on-line
inspection rate of 200 Hz for central PbPb in the TRD and TPC we
can inspect $4 \cdot 10^8$ events per ALICE year of about $2 \cdot
10^6$ effective running seconds. It remains to be shown that
combined TRD-TPC HLT on-line tracking can reduce the selected data
flux down to a level that can be recorded without exceeding the
DAQ bandwidth. With full TRD coverage this promises to cover about
$2 \cdot 10^4$ $\Upsilon$ decays per year, just sufficient for detailed
spectroscopy of the $\Upsilon$ family. 



\subsection{HLT Event Filter for Open Charm Study}

The physics goal of open charm cross section analysis (by D-meson
decay topology/secondary vertex) up to now defines the official
ALICE goals concerning the DAQ bandwidth, taping rate and stored
data size: 10 Hz of central PbPb for $2 \cdot 10^6$ seconds
amounting to $2 \cdot 10^7$ events. If written to tape  we have a
bandwidth of 660 MB/s for the TPC alone and acquire a total of 1.3
Petabytes. Could we get this physics at lower media cost?

The idea is to track the central PbPb events completely at 200 Hz,
on-line with the HLT processors. Having the track information we exclude
the low p$_T$ tracks, which do not contribute to the hadrons emitted at
the D decay vertex. With $<p_T>$ of the D assumed to be 1 GeV/c, the relevant
K$\pi \pi$ (for charged D) and K$\pi$ (for D$^0$) channels result in tracks
above 0.8 GeV/c. If we assume the $<p_T>$ of pions to be 0.6 GeV/c
the fraction of tracks with $p_T > 0.8$ GeV/c is about 0.25\% of the
total. As the secondary D decay vertex is very close to the main
vertex one could also remove large impact parameter non-vertex and
ghost tracks. The procedure would be to define a "pipe" of raw
data along each vertex track at p$_T> $0.8 GeV, which will then be read out.
That procedure needs a further careful simulation. It promises an event
filtering of about 1/10. Within the maximal HLT goal, defined by high E$_T$
jet spectroscopy, that physics represents an HLT activity easily carried out
cooperatively and concurrent to the major task of tracking 200 PbPb events
per second. 

\subsection{Summary of HLT, DAQ and Storage Requirements}

In summary we expect to cover with sufficient statistics per ALICE
running year the observables jet production, and bottonium and
open charm yields, by means of a balanced effort in HLT and DAQ
bandwidths.

\begin{enumerate}
\item HLT is used minimally in first year pp at L=2$\cdot
10^{30}$. We need in principle "only" a vertex-finder algorithm,
but the events (containing 25 superimposed displaced sub-events)
have to be completely cluster-analyzed. At 1000 Hz TPC rate
clustering throughput is about 5.5 GB/s of raw data, but tracking
needs are minimal. Maximum HLT use occurs in the trigger on jets
in central PbPb at 200 Hz TPC. The clustering and tracking effort
rises to up to 20 GB/s.

\item The DAQ bandwidth is intermediate in first year pp minimum bias
runs at 1000 Hz. If the ideal low luminosity of L=10$^{29}$ can be
technically realized, we need no HLT and get about 0.5 $\pm 0.1$
GB/s to DAQ from the TPC. All the other detectors (unless dimuon
adds new tasks) will add 0.1 to 0.2 GB/s. Total DAQ 0.65 $\pm$0.1
GB/s. For higher luminosities we require HLT selection of the
relevant TPC event, thus staying within the same DAQ budget.

\noindent The DAQ bandwidth is maximal for "first year PbPb"
writing 10-20 Hz of central PbPb to tape without any HLT action:
Total DAQ 0.9 $\pm$ 0.2 GB/s. Further "trigger mixes" will
perhaps increase that budget to 1.2 GB/s. The other anticipated
trigger modes of later years will stay at or below the DAQ
bandwidth upon proper trigger mixes under HLT trigger/filter
procedures approaching 200 Hz TPC inspection rate.

\item The storage format is maximal in first year pp because we
shall gather $10^{10}$ min.~bias events in 4-5 months at 1000 Hz
under HLT, which results in about 6.5 Petabytes. Lateron, PbPb
for 2$\cdot 10^6$ seconds per year at a total ALICE rate of 1.2
GB/s will give 2.4 Petabytes.
\end{enumerate}

\subsection{A Further Mode: Data Compression (only) in HLT}

As a remaining option of front end intelligence employment for TPC
data volume reduction let us consider to simply do cluster-finding
on-line and then write to tape without any tracking, trigger
specific search algorithms (vertex, jet, back-to back $e^+e^-$
pair) or invariant mass determinations. Which degree of data
compression can be achieved? Before making estimates we propose,
as a general remark, that in a traditional off-line analysis (like
in NA49 or STAR) one seldom finds reason to return to the raw data
once the analysis chain is set up correctly. However the cluster
level is often returned to . Thus clustered data are almost
equivalent to raw data. The real optimizations, e.g. for specific
ionization, occur at the step from cluster to track.

At the STAR L3 prototype the on-line cluster finder achieves a
sixfold data compression (after zero suppression), which with
about 200.000 clusters per central Au+Au collision leads to a 2 MB
total event size. Thus we saw in STAR that one cluster requires on
average 10 bytes of output. No attempt was made as of yet to bring
down this single cluster format, but there must be considerable
reserve in the 10 bytes. Consider now the present HLT proposal to
replace conventional cluster-finding (in a memory) by "in-flight"
tracklet determination in FPGA based circuits. This way $n$
successive clusters are tied up in a tracklet, say $<n>$ = 7. It
is clear that specification of the tracklet on average will not
require 70 bytes. We might expect that 20 bytes suffice, which would
result in an overall compactification factor of about 15
(and yet could use a re-fitter at the tracklet level or unfolding etc.)
by clustering only. This HLT option would cope with about 150 Hz
central PbPb yet staying compatible with the DAQ bandwidth and
could be an HLT option for later years of ALICE operation. After two years of
off-line operation with the full raw TPC data output, the RHIC STAR
collaboration now converts to record HLT clusters from its TPC
only.

\subsection{Conclusions Concerning Demands on HLT}

From the considerations in all of the above sections it follows
that the maximal envisioned ALICE HLT task will consist in
complete on-line cluster finding and tracking of TPC events, as
motivated by high E$_t$ jet and open charm or bottonium
spectroscopy, at 200 Hz central PbPb as recorded in the TPC. In
all the above cases the by far major computational effort will be
devoted to a complete reconstruction of the event on-line, albeit
with a characteristic lower precision in detail, in comparison to
the ensuing off-line analysis, devoted to the triggered events. In
comparison to the clustering-tracking effort the derivation of
on-line trigger signals, corresponding to specific physics
observables, will require a relatively modest additional
computational effort. The latter computational economy follows
from on-line data handling economy. The track data output does not
get stored, then to be reread as in off-line analysis (highly CPU
time consuming) but is immediately handed, track by track, from
the tracking to the trigger evaluating algorithms.

In ending these  qualitative considerations concerning the
proposed ALICE HLT system let us add a quantitative estimate of
the required HLT CPU power, as based on our experience with the L3
prototype system explored in the STAR experiment at RHIC, BNL. It
is attached there to the TPC subdetector. This TPC has about
one-third of the ADC channels of the ALICE TPC. Furthermore, we
may conservatively estimate that the charged particle multiplicity
should increase, from RHIC $\sqrt{s}$=200 GeV to ALICE
$\sqrt{s}$=5.5 TeV, by a factor of about four. As both TPCs have
the same acceptance we thus expect an ALICE TPC data format of
about 12 times the STAR format. The prototype STAR HLT system
(called L3 there, see \cite{star} was shown in the year 2001 RHIC
running period to cope with the rate of 40 Hz for central AuAu
collisions. For the ALICE HLT we require a 200 Hz rate: a further
factor of 5, bringing the total ALICE computation effort up from
the STAR prototype by a factor of about 60. In STAR 52 PCs of 1997
vintage were employed in the L3 system (mostly $\sim$ 600 MHz).
Adopting expectation from "Moore' law", in extrapolating to ALICE
HLT standard PCs of vintage 2005 we conservatively expect an
overall CPU/network/IO speed increase of about three. This simple
extrapolation results in the expectation that the ALICE HLT
functionality, as envisioned in the previous sections, will
require about 1000 processors. Of course the scaling assumed here
is not trivial and will require advanced methods concerning
networking and computation methods to be outlined in later
sections of the present document.


\section{HLT Trigger for Jets at E$_T\:\ge$ 100 GeV in PbPb}

In this section we describe the application of a cone jet finder
algorithm to derive an on-line HLT jet trigger from TPC inspection
of central PbPb collisions at 200 Hz rate. It is our purpose to
test the jet detection efficiency, the degree of suppression of
accidental background looking like jets, and the specific demand
on CPU time placed by the trigger algorithms over and above the
computing budget already expended in the preceding
clustering-tracking stages.

\begin{figure}[ht]
\begin{center}
\epsfig{figure=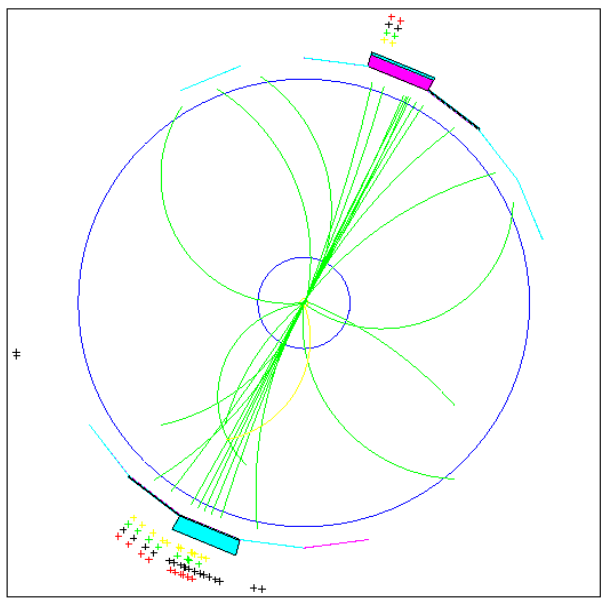,height=8cm, width=8cm}
\caption{A typical di-jet tracking event with E$_T$ of 300 GeV observed by CDF in Tevatron
$p \overline{p}$ collisions \cite{Affolder}.}
\end{center}
\label{fig1}
\end{figure}

The present study is restricted to 100 GeV inclusive jets (without
consideration, at first, of its near back-to-back partner). In
principle the domain of jet total transverse energies that may be
analyzed with ALICE charged particle tracking at 200 Hz PbPb event
rate ranges from 100 GeV (where the statistics is good in one
ALICE run year of about $10^6$ sec) up to about 200 GeV where the
jet statistics has dropped by a factor of about 15 and is, thus,
marginal. The focus at E$_T \: \ge$ 100 GeV represents a
qualitative consideration of the following conditions:

\begin{itemize}
\item The fluctuation in the ratio of charged particle to total jet energy
diminishes with increasing jet charged particle multiplicity
(which is about 10 at 100 GeV).
\item The jet signal should be topologically distinct also in the
high track density of a central PbPb collision, i.e.~have a "high"
charged track multiplicity to avoid observational bias.
\item The leading jet particle energy should not exceed the limits of
accurate enough  $p_T$ measurement in the ITS-TPC-TRD complex,
i.e.~$p_T \le 50$ GeV/c.
\item On the other hand most of the total $E_T$ should be in
particles with $p_T \ge $ 2 GeV/c.
\end{itemize}

\begin{figure}[ht]
\begin{center}
\epsfig{figure=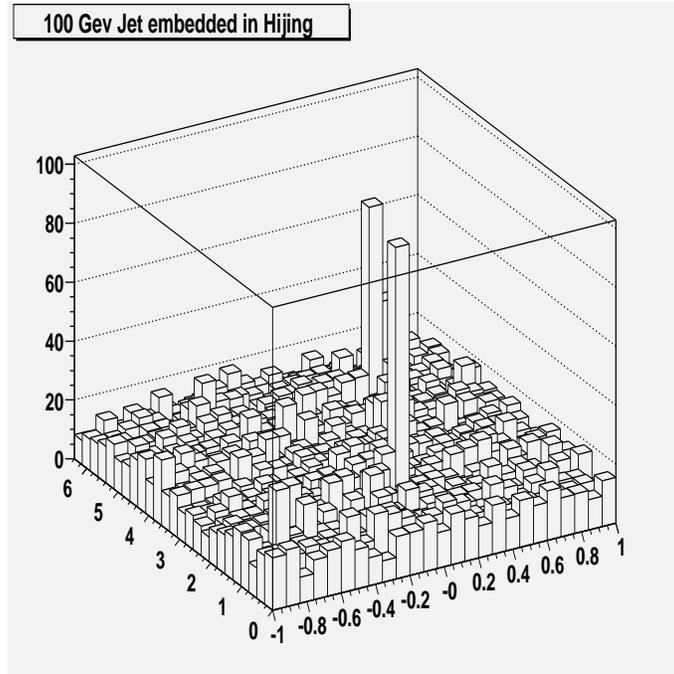,height=9cm,width=9cm}
\caption{Typical 100 GeV Di-jet event with granularity $\Delta \eta = 0.1$ and $\Delta \phi = 0.25$.}
\end{center}
\label{fig2}
\end{figure}

\noindent In Fig. \ref{fig1} we illustrate a typical di-jet
tracking event with E$_T$ of 300 GeV observed by CDF in Tevatron
$p \overline{p}$ collisions \cite{Affolder}.  The jet related
topological features dominate, by far, the non-jet related track
background. This will be similar in pp at ALICE. However, note
that the jet cone algorithm \cite{ConeAlgo} employed in $p
\overline{p}$ is aiming at an exhaustive coverage of the jet track
manifold, down to tracks at rather large angle relative to the jet
axis (to reconstruct the full fragmentation function). One thus
normally employs a jet cone radius of $R=0.7$ in the
plane of pseudorapidity $\eta$ vs. azimuthal angle $\phi$ (ranging
from zero to 2 $\pi$). In ALICE this would comprise almost the
full rapidity acceptance, and 22\% of the azimuthal range. This
may be appropriate for ALICE pp jet study but is clearly not
useful in a straight forward manner under LHC PbPb conditions,
because every such cone near mid-rapidity would now contain
charged tracks with a total of about 0.7 TeV transverse energy,
with a fluctuation RMS of about 40 GeV. This follows from a HIJING
simulation assuming a midrapidity charged particle density of
6000. As the average charged track total E$_T$ of a 100 GeV jet is
60 GeV \cite{Affolder}, the jets can not be well disentangled. Let
us emphasize, again, that the HLT task considered here is only to
find the jet candidate events. This requires a narrow jet cone
finder algorithm - to exploit the overall typological jet pattern.
Off-line analysis will subsequently study the jet activities in
any cone required.

We wish to stay with the cone algorithm for its relative
computational efficiency (required, at least, in the on-line
analysis) but consider a significant reduction of R, to 0.3 or
0.2, for the process of on-line trigger generation, which is
{\bf merely jet finding}. The triggered events will then be written to
storage in full raw data format, for off-line jet analysis of any
kind. The jet finding conditions are illustrated in Fig. \ref{fig2}. Here
and in the following, we use the code Pythia version 6.161 to generate
elementary pp events with a contained hard parton scattering
creating scattered partons of 100 GeV transverse momentum. These
elementary events are then analyzed with the standard cone algorithm 
(with $R=0.7$) to identify the highest energy jet in the event, representing the
outcome of the initial parton scattering. The charged tracks found
within the cone of this particular jet are then imbedded into a
central PbPb HIJING event, represented by the charged track
distribution in $\eta$ and $\phi$. Actually it turns out that the
HIJING average track and energy density is flat within the entire
ALICE TPC acceptance of $\mid \eta_{CM}\mid  \le \:1$. Fig \ref{fig2} thus
represents the image of a typical 100 GeV di-jet event in a Lego
plot with granularity $\Delta \eta = 0.1$ and $\Delta \phi = 0.25$.
The distribution and fluctuation of charged track transverse
energy, observed here, would be typical of calorimetric summative
$E_T$ analysis. In this picture the typical jet correlation of
high E$_T$ tracks, closely packed in a narrow cone about the jet
axis, creates a topologically distinct pattern that stands out
well above the background.

In ALICE we have to base recognition of the topological jet
signature on an appropriate analysis of jet cone correlation among
high E$_T$ individual charged particle tracks.  For $p
\overline{p}$ collisions at the Tevatron the CDF Collaboration has
recently published a comprehensive study of jet physics, based on
charged particle tracking only \cite {Affolder2}. They study
the systematic evolution of jet fragmentation functions upon
variation of cone jet-finder algorithm, downwards from R=0.7 to
0.2 and note, in particular, that at R=0.2 still 80\% of the total
charged particle $E_T$ is contained in the jet cone. This finding
encourages us  to work with cone radii of 0.3 and 0.2,
respectively, for on-line jet finding to result in a fast
trigger, in PbPb central collisions where higher cone radii would
meet with increasing background fluctuations.

Within an on-line HLT clustering-tracking procedure for the entire
event each track emerges with a determined center of mass
momentum vector. For CPU economy of the ensuing jet finder algorithm it is essential to
select the relevant track candidates right then, rather than
depositing all tracks in a register that the cone finder would
have to re-read. We thus base the jet finder on a cone correlation
of high $p_T$ tracks which are handed, above a certain $p_T$
cutoff, directly from tracking to the jet finder algorithm. Within
the latter we then inspect the event in terms of requiring n
charged tracks above a $p_T$ or E$_T$ cutoff  of m GeV, contained
within a cone of radius R=0.2 or 0.3 in $\eta$ and $\phi$.

\begin{figure}[ht]
\begin{center}
\epsfig{figure=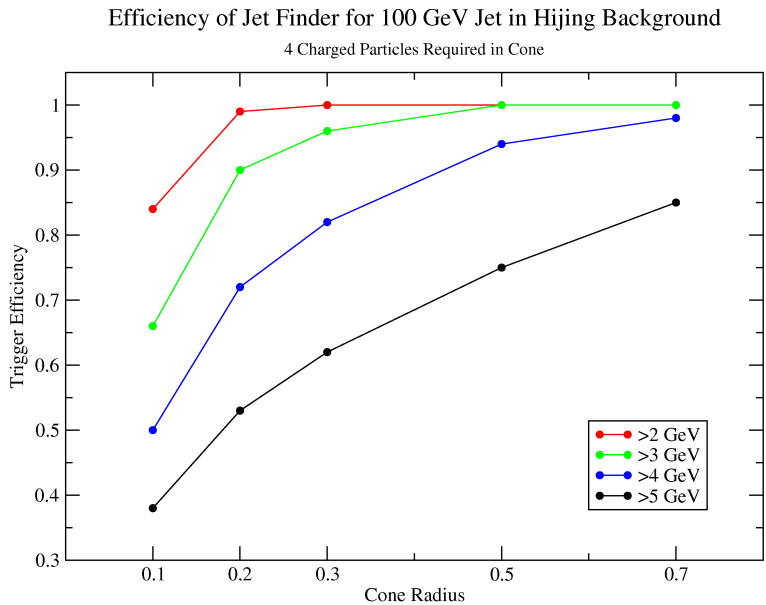,width=12cm} 
\caption{Efficiency of
cone jet finder trigger applied for different cone radii and
threshold. At least 4 charged particles over threshold are
required to be inside the cone.}
\end{center}
\label{fig3}
\end{figure}

In the simulation with elementary Pythia jet identified events of
100 GeV initial partonic transverse energy/momentum, imbedded into
HIJING simulated events for central PbPb collisions, at ALICE
energy and within the ALICE acceptance, we have studied the
jet-detection efficiency and accidental background rate of various
trigger-defining options, employed in the cone jet-finder
algorithm. As an example Fig. 3  shows the resulting jet
recognition efficiency requiring at least four charged tracks
correlated within jet cone radii ranging from 0.1 to 0.7, for
various track $p_T$ cuts ranging from above 2 GeV/c to above 5
GeV/c. Fig. 4  illustrates the background trigger rate of these cone trigger
options, in terms of accidental background being created by random
$p_T, \: \phi$ fluctuations in average HIJING central PbPb
simulated events. The finite ALICE transversal momentum resolution
should not lead to significant changes.

\begin{figure}[ht]
\begin{center}
\epsfig{figure=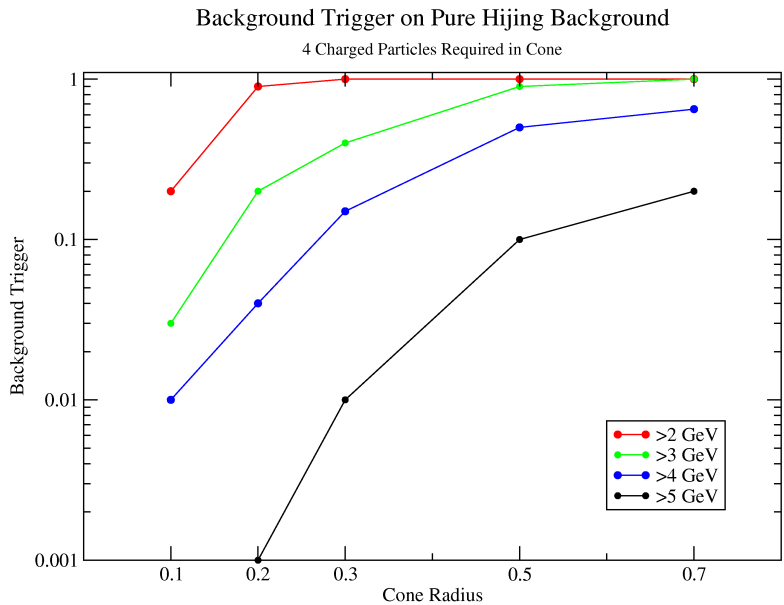,width=12cm}
\caption{Background trigger rate of cone jet finder
trigger measured on pure Hijing background events. Parameters are
the same as in Fig. 3.}
\end{center}
\label{fig4}
\end{figure}

At the level of this preliminary study an optimum of trigger
efficiency vs. accidental trigger background rate may be
accomplished in requiring 4 charged tracks above 4 GeV transverse
momentum within a jet cone of R=0.2.  The resulting jet
efficiency, of about 0.72, coincides with a background trigger 
rate of about 4 \%. I.e.~this trigger mode, as applied in
HLT central PbPb collisions inspection at 200 Hz, created a jet
candidate trigger rate of about 8 events per second that can
easily be written to tape, for comprehensive off-line analysis.
Furthermore, the events collected under this HLT trigger mode can
be considered as almost bias free concerning any other physics
observable of interest: more than 99\% of the resulting HLT
triggers are based on random event-by-event fluctuations in the
high $p_T$ sector, immaterial to many other ALICE physics
observables.

We thus argue that an appropriate on-line HLT jet trigger based on
charged track 3-momentum determination on-line, at the tracking
stage, and on an optimized cone-type jet-finder algorithm, will
offer the required jet recognition efficiency, within a
selectivity above background that will reduce the 200 Hz rate of
HLT inspected central PbPb TPC events, down to a candidate trigger
rate of about 8 Hz (essentially bias free as concerns analysis of
any other observables, except for high E$_T$ jets). This latter
event rate fits well within the overall anticipated ALICE TPC to
DAQ bandwidth, of 10-20 events per second. Thus it will be
possible to record other trigger modes concurrently.

It remains to be shown that the HLT jet-cone trigger search
algorithm, as implied in this section, does not inflict a
significant additional budget concerning CPU time, in addition to
the -already maximal- HLT task, to perform cluster and track
analysis for central PbPb TPC events at 200 Hz rate. At present we
estimate the jet cone algorithm to require less than one second in
the mode illustrated above. Both this estimate, and also the
accidental background should improve with a further, more
comprehensive optimization of the detailed trigger conditions.

A higher jet recognition efficiency might
actually result from exploiting the approximate back-to-back
topology of dijet production (this trigger mode could also be more
interesting, physics-wise!). One could thus search with a double
cone algorithm, relaxing the charged high p$_T$ track number
requirement in each cone from the above 4-5 down to 2-3. In a
single cone this leads to jet recognition with up to 90\%
efficiency but creates too high a background accidental rate. The
additional topological constraint implied by di-jet events should
significantly reduce the background rate, yet leaving one with an
overall dijet efficiency of about 80\%.


In summary we expect, from the present level of jet physics
simulation, to be able to present an optimized HLT cone jet finder
trigger for inclusive jets of $E_T \ge$ 100 GeV that delivers
75-80 \% efficiency with a background trigger rate of
1-2 \%. The expected CPU time budget is below 20\% of the effort
required for cluster finding and tracking.


\end{document}